\documentclass[twocolumn]{aastex631}

\usepackage{IEEEtrantools} 
\usepackage{amsmath}

\newcolumntype{L}[1]{>{\raggedright\arraybackslash}p{#1}}
\newcolumntype{C}[1]{>{\centering\arraybackslash}p{#1}}
\newcolumntype{R}[1]{>{\raggedleft\arraybackslash}p{#1}}

\newcommand{\msun}{\ensuremath{M_{\odot}}}
\newcommand{\lsun}{\ensuremath{L_{\odot}}}
\newcommand{\rsun}{\ensuremath{R_{\odot}}}
\newcommand{\rhosun}{\ensuremath{\rho_{\odot}}}
\newcommand{\numax}{\ensuremath{\nu_{\textrm{max}}}}
\newcommand{\dnu}{\ensuremath{\Delta\nu}}

\newcommand{\teff}{\ensuremath{T_{\textrm{eff}}\:}}
\newcommand{\logg}{\mbox{$\log g$}}
\newcommand{\fbol}{\mbox{$f_{\rm{bol}}$}}

\newcommand{\muHz}{\mbox{$\mu$Hz}}
\newcommand{\feh}{\mbox{$\rm{[Fe/H]}$}}

\newcommand{\afeh}{\mbox{$\rm{[\alpha/Fe]}$}}

\newcommand{\rhostar}{\mbox{$\rho_{\star}$}}
\newcommand{\rstar}{\mbox{$R_{\star}$}}
\newcommand{\lstar}{\mbox{$L_{\star}$}}
\newcommand{\mstar}{\mbox{$M_{\star}$}}

\newcommand{\target}{KIC\,8144907}
\newcommand{\met}{\mbox{$-2.66 \pm 0.08$}}
\newcommand{\alp}{\mbox{$0.38 \pm 0.06$}}

\newcommand{\mass}{\mbox{$0.79\pm0.02 \rm{(ran)} \pm 0.01 \rm{(sys)}$}}
\newcommand{\age}{\mbox{$12.0\pm0.6 \rm{(ran)} \pm 0.4 \rm{(sys)}$}}
\newcommand{\radius}{\mbox{$3.62\pm0.04 \rm{(ran)} \pm 0.02 \rm{(sys)}$}}
\newcommand{\density}{\mbox{$2.37\pm0.02 \rm{(ran)} \pm 0.04 \rm{(sys)}$}}
\newcommand{\loggval}{\mbox{$3.221\pm0.002 \rm{(ran)} \pm 0.006 \rm{(sys)}$}}
\newcommand{\lum}{\mbox{$10.42\pm0.43 \rm{(ran)} \pm 0.81 \rm{(sys)}$}}

\newcommand{\kep}{\mbox{\textit{Kepler}}}

\newcommand{\gaia}{\mbox{\textit{Gaia}}}

\usepackage{xcolor}

\newcommand{\rev}[1]{\textcolor{black}{#1}}

\definecolor{meridithgreen}{RGB}{0, 150, 0}

\shorttitle{Asteroseismology of a Very Metal-Poor Star}  
\shortauthors{Huber et al.}

\begin{document}

\title{Stellar Models are Reliable at Low Metallicity: An Asteroseismic Age for the Ancient Very Metal-Poor Star KIC\,8144907} 

\suppressAffiliations

\correspondingauthor{Daniel Huber}
\email{huberd@hawaii.edu}

\author[0000-0001-8832-4488]{Daniel Huber}
\affiliation{Institute for Astronomy, University of Hawai`i, 2680 Woodlawn Drive, Honolulu, HI 96822, USA}
\affiliation{Sydney Institute for Astronomy (SIfA), School of Physics, University of Sydney, NSW 2006, Australia}

\author[0000-0003-4538-9518]{Ditte Slumstrup}
\affiliation{European Southern Observatory, Alonso de Cordova 3107, Vitacura, Chile}
\affiliation{Instituto de Estudios Astrof\'isicos, Facultad de Ingenier\'ia y Ciencias, Universidad Diego Portales, Av. Ej\'ercito Libertador 441, Santiago, Chile}

\author[0000-0003-2400-6960]{Marc Hon}
\affiliation{Kavli Institute for Astrophysics and Space Research, Massachusetts Institute of Technology,
77 Massachusetts Avenue, Cambridge, MA 02139, USA}
\affiliation{Institute for Astronomy, University of Hawai`i, 2680 Woodlawn Drive, Honolulu, HI 96822, USA}

\author[0000-0003-3020-4437]{Yaguang Li}
\affiliation{Institute for Astronomy, University of Hawai`i, 2680 Woodlawn Drive, Honolulu, HI 96822, USA}

\author{Victor Aguirre B{\o}rsen-Koch}
\affiliation{DARK, Niels Bohr Institute, University of Copenhagen, Jagtvej 128,18, 2200, Copenhagen, Denmark}

\author[0000-0001-5222-4661]{Timothy R. Bedding}
\affiliation{Sydney Institute for Astronomy (SIfA), School of Physics, University of Sydney, NSW 2006, Australia}

\author[0000-0002-8717-127X]{Meridith Joyce}
\affiliation{Konkoly Observatory, HUN-REN Research Centre for Astronomy and Earth Sciences, Konkoly-Thege Mikl\'os \'ut 15-17, H-1121, Budapest, Hungary}
\affiliation{CSFK, MTA Centre of Excellence, Budapest, Konkoly-Thege Mikl\'os \'ut 15-17, H-1121, Budapest, Hungary}

\author{J.\ M.\ Joel Ong}
\affiliation{Institute for Astronomy, University of Hawai`i, 2680 Woodlawn Drive, Honolulu, HI 96822, USA}

\author{Aldo Serenelli}
\affiliation{Institute of Space Sciences (ICE, CSIC), Carrer de Can Magrans S/N, Campus UAB, Cerdanyola del Valles, E-08193, Spain}
\affiliation{Institut d'Estudis Espacials de Catalunya (IEEC), 08860 Castelldefels (Barcelona), Spain}

\author[0000-0002-4879-3519]{Dennis Stello}
\affiliation{School of Physics, University of New South Wales, NSW 2052, Australia}
\affiliation{Sydney Institute for Astronomy (SIfA), School of Physics, University of Sydney, NSW 2006, Australia}
\affiliation{ARC Centre of Excellence for All Sky Astrophysics in Three Dimensions (ASTRO-3D)}

\author[0000-0002-2580-3614]{Travis Berger}
\affiliation{Space Telescope Science Institute, 3700 San Martin Drive, Baltimore, MD 21218, USA}

\author[0000-0003-4976-9980]{Samuel K. Grunblatt}
\affiliation{William H. Miller III Department of Physics and Astronomy, Johns Hopkins University, 3400 N Charles St, Baltimore, MD 21218, USA}

\author[0000-0002-0371-1647]{Michael Greklek-McKeon}
\affiliation{Division of Geological and Planetary Sciences, California Institute of Technology, Pasadena, CA, 91125, USA}

\author[0000-0003-3618-7535]{Teruyuki Hirano}
\affiliation{Astrobiology Center, 2-21-1 Osawa, Mitaka, Tokyo 181-8588, Japan}
\affiliation{National Astronomical Observatory of Japan, 2-21-1 Osawa, Mitaka, Tokyo 181-8588, Japan}

\author[0000-0001-6196-5162]{Evan N. Kirby}
\affiliation{Department of Physics and Astronomy, University of Notre Dame, 225 Nieuwland Science Hall, Notre Dame, IN 46556, USA}

\author[0000-0002-7549-7766]{Marc H. Pinsonneault}
\affiliation{Department of Astronomy, The Ohio State University, Columbus, OH 43210, USA}

\author{Arthur Alencastro Puls}
\affiliation{Goethe University Frankfurt, Institute for Applied Physics, Max-von-Laue-Str. 12, 60438, Frankfurt am Main, Germany}

\author[0000-0002-7550-7151]{Joel Zinn}
\affiliation{Department of Physics and Astronomy, California State University, Long Beach, Long Beach, CA 90840, USA}

\begin{abstract}
Very metal-poor stars ($\feh <-2$) are important laboratories for testing stellar models and reconstructing the formation history of our galaxy. Asteroseismology is a powerful tool to probe stellar interiors and measure ages, but few asteroseismic detections are known in very metal-poor stars and none have allowed detailed modeling of oscillation frequencies. We report the discovery of a low-luminosity \kep\ red giant (\target) with  high S/N oscillations, $\feh = \met$ and $\afeh= \alp$, making it by far the most metal-poor star to date for which detailed asteroseismic modeling is possible. By combining the oscillation spectrum from \kep\ with high-resolution spectroscopy we measure an asteroseismic mass and age of \mass\,\msun\ and \age\,Gyr, with remarkable agreement across different codes and input physics, demonstrating that stellar models and asteroseismology are reliable for very metal-poor stars when individual frequencies are used. The results also provide a direct age anchor for the early formation of the Milky Way, implying that substantial star formation did not commence until redshift z$\approx$3 (\rev{if the star formed in-situ}) or that the Milky Way has undergone merger events for at least $\approx$\,12 Gyr (if the star was accreted by a dwarf satellite merger such as Gaia Enceladus).
\end{abstract}


\section{Introduction}

The study of the chemo-dynamical history of stellar populations in our galaxy is a rapidly evolving field \citep[see e.g.][for recent reviews]{bland-hawthorn_galaxy_2016,helmi_streams_2020,bonaca_stellar_2024}. A particularly important population in our galaxy are metal-poor stars in the Galactic halo, with theories predicting that they formed either through a combination of collapse of gas during the early stages of galaxy formation \citep{eggen_evidence_1962}, the accretion of satellite galaxies at later times \citep{searle_composition_1978}, a combination of early and late accretion of satellite galaxies \citep{bullock_tracing_2005} or ejections of Galactic disc/bulge stars via dynamical interactions \citep[e.g.][]{purcell_heated_2010}. 

A key element to advance our understanding of halo stars is stellar ages. The combination of red giant asteroseismology from space-based missions such as \kep, K2 and TESS with spectroscopic surveys such as APOGEE \citep{holtzman_abundances_2015}, Gaia-ESO \citep{Gilmore2012}, GALAH \citep{de_silva_galah_2015} and LAMOST \citep{cui_large_2012} has provided systematic age constraints for thousands of field red giants \citep[e.g.][]{miglio_probing_2009,anders_galactic_2017,stello_k2_2017,sharma_k2-hermes_2019,grunblatt_age-dating_2021,borre_age_2022,alencastro_puls_chemo-dynamics_2022,willett_evolution_2023,schonhut-stasik++2024-apok2} and enabled constraints on mass-loss for metal-poor populations in globular clusters \citep{miglio_detection_2016, howell_integrated_2022, tailo_asteroseismology_2022, howell_first_2024}. However, all these constraints rely on asteroseismic scaling relations, which are poorly calibrated for metal-poor stars, show systematic offsets compared to expected astrophysical priors for halo stars \citep{epstein_testing_2014,matsuno_star_2021}, and yield high random age uncertainties \citep{moser_uncertainties_2023}.

The gold-standard for ages of field stars comes from asteroseismic modeling of individual oscillation frequencies. This ``boutique'' modeling is often restricted to main-sequence and subgiant stars due to the fast evolution of red giants, which requires more intensive computation. The issue of surface correction also has to be treated carefully \citep{kjeldsen_correcting_2008, ball_surface_2018,ong_mixed_2021,li_prescription_2023}. 
Because oscillation amplitudes are very small for low-luminosity stars \citep{Kjeldsen1995, huber_solar-like_2011, mosser_characterization_2012, kallinger_connection_2014}, frequency modeling has been applied only to two stars with $\feh < -1$: the nearby subgiant $\nu$\,Indi ($\feh = -1.5$\,dex), an ``in-situ'' halo star \citep{chaplin_age_2020}, and KIC7341231 ($\feh = -1.6$\,dex), which was used to constrain an interior rotation profile \citep{deheuvels_seismic_2012}.  

Confronting state-of-the-art models with detailed observations at low metallicities has ramifications for a wide range subfields of astrophysics that use stellar models. Asteroseismic frequency modeling of metal-poor stars is important to test these stellar models. For example, prescriptions for convective energy transport in stellar interiors \citep[e.g.,][]{trampedach_improvements_2014,magic_stagger-grid_2015,Tayar2017, JoyceChaboyer2018NotAllStars, joyce_review_2023}. 
The combination of asteroseismic constraints with effective temperatures and chemical abundances also test the significance of non-local thermodynamic equilibrium (non-LTE) effects in stellar atmospheres \citep[e.g.,][]{bergemann_non-lte_2012}, which are expected to depend on metallicity.

Here, we present the results of a survey for metal-poor stars among the \kep\  oscillating red giant sample. We have found 16 oscillating stars with $\feh < -1$ and present the first detailed asteroseismic modeling of a very-metal poor star ($\feh < -2$).

\section{Spectroscopy}

\subsection{Survey for Very Metal-Poor Stars}

We performed a spectroscopic survey for very metal-poor stars among known oscillating \kep\ red giants. Targets were selected from a catalog of oscillating giants with disk height $z>\,3$\,kpc \citep{mathur_probing_2016}, and by kinematically selecting likely halo stars using Gaia. For the latter, we calculated tangential motions and used a synthetic population from Galaxia \citep{sharma_galaxia_2011} to define a fiducial locus of low-luminosity red giants with kinematics that are consistent with halo stars.

Observations were performed between 2017 and 2019 with the Supernova Integrated Field Spectrograph (SNIFS) mounted on UH88 telescope \citep{lantz_snifs_2004}, Keck/HIRES \citep{vogt_hires_1994} and Subaru/HDS \citep{noguchi_high_2002}. We performed initial low-resolution spectroscopy with SNIFS (R=3000) of $\approx$\,50 stars using template spectra to select candidate very metal-poor stars, which were then followed up with high-resolution spectroscopy. High-resolution spectra were obtained with a typical SNR of 20-40 per pixel, depending on target brightness. An initial abundance analysis was performed using iSpec \citep{blanco-cuaresma_determining_2014} with the \verb|spectrum| synthesis code using the wavelength range 600--680\,nm. We identified 16 targets as metal-poor ($\feh < -1$; see Appendix). \target\ (2MASS J18485977+4401183, $Ks=11.1$\,mag) was identified as a high-priority follow-up target given its rich asteroseismic data from \kep\ and very low metallicity inferred from a HIRES spectrum obtained in good conditions on 2019 June 25 with a 300~s exposure and a 1.15'' wide slit. We obtained follow-up spectra of \target\ in 2020 using Subaru/HDS with the I2a setup (500--760\,nm) and the default 0.4'' slit, resulting in a spectral resolution of $R\approx90000$. We co-added spectra with a total integration time of 1.2 hours, resulting in an SNR of 70 between 565--585\,nm.

\subsection{Atmospheric Parameters and Abundances}

\begin{figure}
\begin{center}
\resizebox{\hsize}{!}{\includegraphics{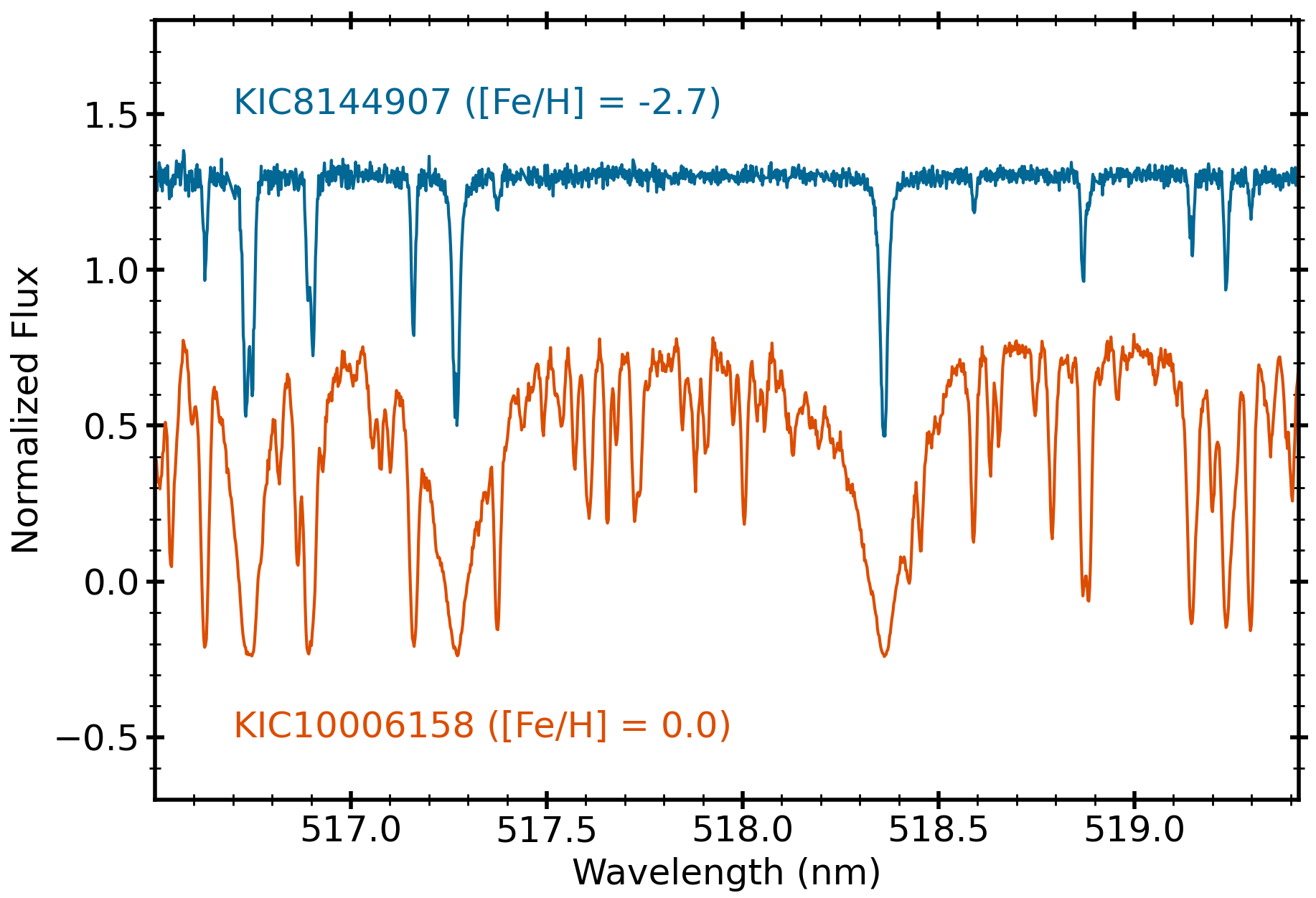}}
\caption{Subaru/HDS spectrum of \target\ near the Mg triplet. The red line shows a Keck/HIRES spectrum of the solar-metallicity star KIC\,10006158 for comparison.}
\label{fig:spectrum}
\end{center}
\end{figure}

\begin{figure*}
\begin{center}
\resizebox{\hsize}{!}{\includegraphics{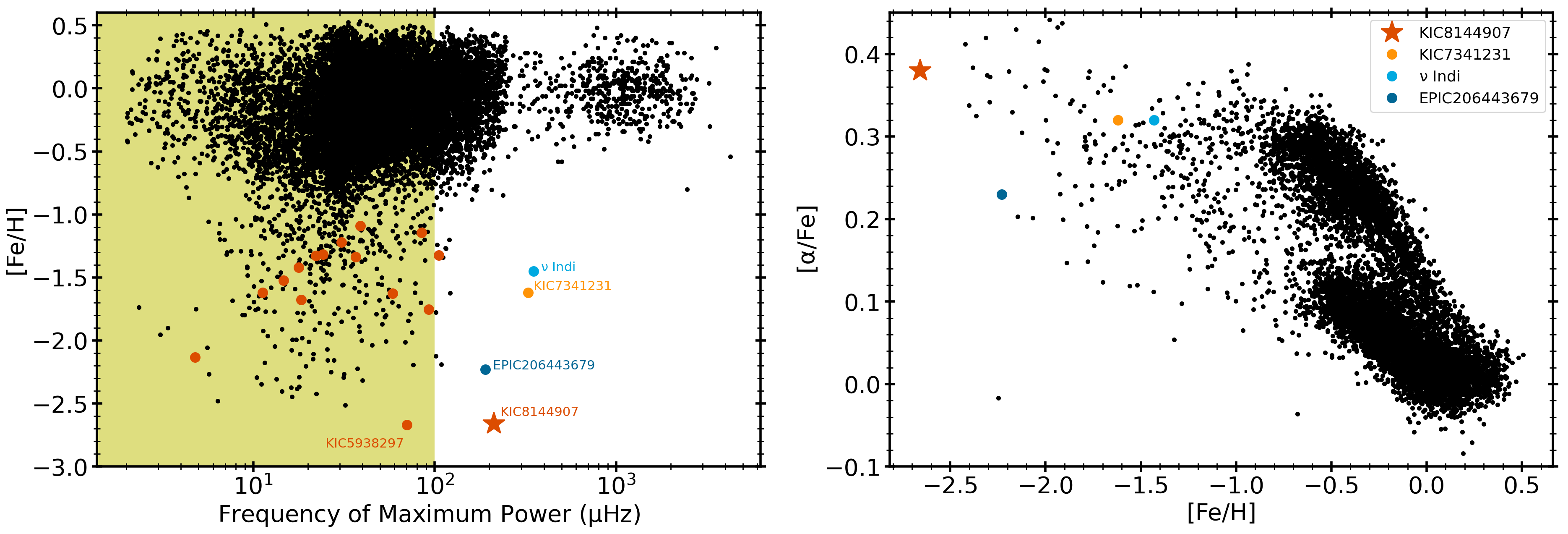}}
\caption{Left: Iron abundance versus frequency of maximum power for stars with asteroseismic detections. Literature values (black points) are from \citet{pinsonneault_apokasc_2014}, \citet{serenelli_first_2017}, \citet{matsuno_star_2021} and \citet{schonhut-stasik++2024-apok2}. Individual literature values are from \citet{chaplin_age_2020} ($\nu$\,Indi), \citet{valentini_masses_2019} (EPIC\,206443679), and \citet{deheuvels_seismic_2014} (KIC7341231). New data from this study are shown as red symbols. The shaded area marks stars generally not amenable to frequency modeling ($\numax<100$\,\muHz). Right: Same as the left panel but showing $\alpha$ abundance versus metallicity.}
\label{fig:sample_spec}
\end{center}
\end{figure*}

Figure \ref{fig:spectrum} shows a portion of our combined optical spectrum for \target\ compared to a solar metallicity star. We used equivalent width measurements to derive atmospheric parameters and chemical abundances, requiring excitation and ionization equilibrium. We varied the temperature until no trend was seen between [Fe/H] and excitation potential while fixing surface gravity to 3.26, as measured from asteroseismic scaling relations \citep{yu_asteroseismology_2018}. 
We ensured agreement between FeI and FeII absorption lines was within one standard deviation. The equivalent widths were measured with DAOSPEC \citep{stetson2008} and the individual absorption line abundances were calculated with the auxiliary program \textit{Abundance} with SPECTRUM in LTE \citep{Gray1994}. The chosen absorption lines and accompanying oscillator strengths are from \citet{slumstrup2019} for our wavelength range, which is 495--675 nm with a few minor gaps. The solar abundances are those of \citet{grevesse1998} and the stellar atmosphere models are ATLAS9 \citep{castelli_new_2004}. Due to the very low metallicity we applied non-LTE departure coefficients from the INSPECT database version 1.0\footnote{available at \url{www.inspect-stars.com}} \citep{Bergemann2012, Lind2012} to derive final atmospheric parameters of $\teff = 5400\pm200$\,K, $\feh = -2.66\pm0.08$\,dex and $\afeh = 0.38\pm0.06$\,dex.
Uncertainties were calculated by varying each parameter until at least a 3$\sigma$ uncertainty is produced on the slope of [Fe/H] vs. excitation potential. We did not detect signs of $r$-process enhancement, such as Eu (a signature for accreted halo stars; \citealt{matsuno_r_2021}).

Figure \ref{fig:sample_spec} compares \target\ to known oscillating stars in abundance and frequency space. \target\ is the most metal-poor oscillating star that is not an evolved red-giant star ($\numax > 100$\,\muHz) and the most metal-poor oscillating star with measured $\alpha$ abundances. EPIC\,206443679 \citep{valentini_masses_2019} is a very metal-poor star with similar \numax, but was only observed for 70 days with K2, limiting the SNR and frequency resolution of the asteroseismic detection. 
KIC\,5938297 is another very metal-poor but more evolved star ($\numax \approx 80$\muHz) identified by our survey.

\section{Asteroseismology}

\subsection{Oscillation Frequencies}

\begin{figure*}
\begin{center}
\resizebox{\hsize}{!}{\includegraphics{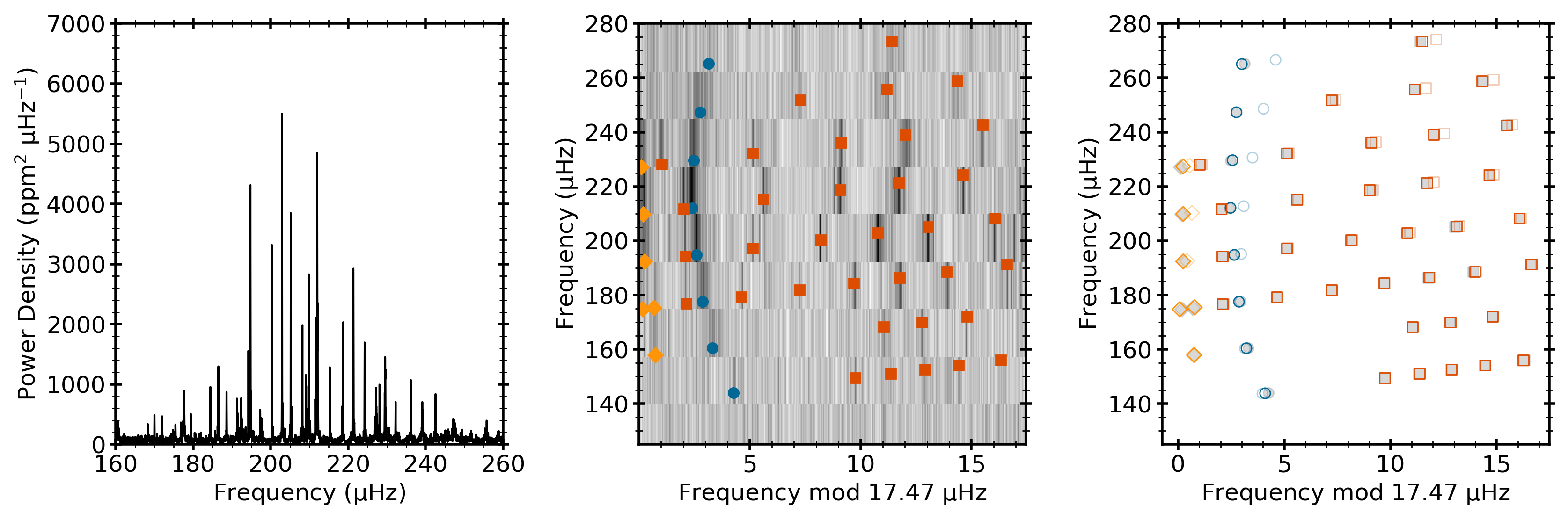}}
\caption{\textit{Left:} Power spectrum of the 4-year \kep\ timeseries of \target\ centered on the oscillations. \textit{Middle:} \'{E}chelle diagram (greyscale) and extracted frequencies for radial (circles), dipole (squares) and quadrupole (diamonds) modes. \textit{Right}: \'{E}chelle diagram showing frequencies from the best fitting model from BeSPP (Section 3.3.1) before (faint open symbols) and after (open symbols) applying a surface correction. Grey symbols show the observed frequencies.}
\label{fig:seismo_obs}
\end{center}
\end{figure*}

\target\ was observed by \kep\ for its entire mission using long-cadence mode. We used the PDC-SAP light curve \citep{stumpe_kepler_2012,smith_kepler_2012} and computed the power spectrum, which shows solar-like oscillations with high SNR (Fig.\ \ref{fig:seismo_obs}, left panel). To extract frequencies, we smoothed the spectrum to a resolution of 0.1\,\muHz\ and measured the positions of the peaks down to a signal-to-noise ratio of 3.5. We identified 49 mode frequencies, which are listed in Table \ref{tab:freqs}. 

The left and middle panel of Figure \ref{fig:seismo_obs} show the power spectrum and \'{e}chelle diagram. Note that the peaks at the top of the \'{e}chelle diagram are aliases reflected around the Nyquist frequency (283\,\muHz). The data show clear ridges of radial and quadrupole modes, plus a rich series of mixed dipole modes that are consistent with a star at the base of the RGB.  \citet{kuszlewicz_mixed-mode_2023} showed that \target\ has a strongly enhanced mixed-mode coupling strength and rotational splittings consistent with the core rotation rates of the general \kep\ sample.

\begin{table}
\begin{center}
\caption{Oscillation frequencies and mode identifications for \target. We adopt twice the frequency resolution (0.02\muHz) as a conservative uncertainty estimate for each frequency.}
\vspace{0.1cm}
\begin{tabular}{c c | c c | c c}        
$f(\muHz)$ & $l$ & $f(\muHz)$ & $l$ & $f(\muHz)$ & $l$  \\
\hline 
144.02	&	0	&	184.40	&	1	&	221.36	&	1 \\
149.51	&	1	&	186.47	&	1	&	224.25	&	1 \\
151.11	&	1	&	188.59	&	1	&	227.22	&	2 \\
152.66	&	1	&	191.30	&	1	&	228.15	&	1 \\
154.19	&	1	&	192.42	&	2	&	229.59	&	0 \\
156.07	&	1	&	194.27	&	1	&	232.25	&	1 \\
157.98	&	2	&	194.78	&	0	&	236.23	&	1 \\
160.54	&	0	&	197.31	&	1	&	239.13	&	1 \\
168.27	&	1	&	200.36	&	1	&	242.61	&	1 \\
170.01	&	1	&	202.94	&	1	&	247.35	&	0 \\
172.03	&	1	&	205.21	&	1	&	251.87	&	1 \\
174.88	&	2	&	208.23	&	1	&	255.75  &	1 \\
175.39	&	2	&	209.83	&	2	&	258.94	&	1 \\
176.84	&	1	&	211.66	&	1	&	265.19	&	0 \\
177.57	&	0	&	212.04	&	0	&	273.45	&	1 \\
179.33	&	1	&	215.25	&	1	&			&  ---   \\
181.94	&	1	&	218.71	&	1	&			&  ---  \\
\hline         
\end{tabular}
\label{tab:freqs}
\end{center}
\end{table}

\subsection{Bolometric Flux and Luminosity}

To calculate the bolometric flux (\fbol) we used 2MASS K-band photometry \citep{skrutskie_two_2006} and bolometric corrections from MIST isochrones \citep{choi_mesa_2016}, as implemented in \texttt{isoclassify} \citep{huber_asteroseismology_2017}, interpolated to the spectroscopic \teff, \feh\ and asteroseismic \logg. Interstellar extinction was calculated using the 3D map by \citet{green_three-dimensional_2015}, yielding Av=0.15\,mag. We assumed uncertainties of 0.03\,mag on the bolometric and extinction correction \citep{tayar_guide_2022}. We find $\fbol = 2.02 \pm 0.06 \times 10^{-10}$ erg\,s$^{-1}$\,cm$^{-2}$, which combined with a distance of $1326\pm20$\,pc from the \gaia\ DR3 parallax \citep{gaia_collaboration_gaia_2021,lindegren_gaia_2021} yields a luminosity of $11.13\pm0.45$\,\lsun.

\subsection{Frequency Modeling}

We applied six independent modeling efforts to constrain potential systematic errors from different input physics and modeling codes \citep[e.g.][]{chaplin_asteroseismic_2014,silva_aguirre_ages_2015}.

\subsubsection{BeSPP}

A.S.\ computed models using Garstec \citep{Weiss:2008} with input physics following \citet{serenelli_first_2017}. In summary, nuclear reaction rates are from Solar Fusion II \citep{adelberger:2011}, the equation of state is FreeEOS \citep{freeeos:2012}, atomic opacities are from OPAL \citep{opal:1996}, and low-temperature opacities are from the Wichita group \citep{ferguson_low-temperature_2005}. Stellar atmospheres are modeled using a VAL-C T-$\tau$ relation from \citet{vernazza:1981}. Mixing length theory is implemented following \citep{cox_principles_1968}. The mixing length parameter $\alpha_{\rm MLT}=1.9980$ is determined consistently by a solar model calibration. Chemical mixing is treated as a diffusive process. Overshooting is modeled following the exponential decay model of \citet{freytag_hydrodynamical_1996} and a restriction for small convective cores as described by \citet{magic:2010}. Gravitational settling is included in the models. In addition, extra-mixing below the convective envelope is included following the parametrization by \citet{vandenberg:2012} with a metallicity dependent calibration of the diffusion coefficient that reproduces the lithium depletion in the Sun and the moderate depletion of metals in old globular clusters. 

The model grid covered a mass range between 0.6 and 1.2~\msun\ with a step of 0.01~\msun, and [Fe/H] between $-1.0$ and $-3.0$. All models have been computed with [$\alpha$/Fe]= 0.4. Evolutionary tracks ran from the Hayashi track to the low-luminosity red giant branch, in practice stopping when the surface gravity reached $\logg=2.8$. Along each track, about 2000 stellar models were computed. Stellar tracks are pruned to retain only models that are close enough to the surface parameters of the star (within 3$\sigma$ for \teff and 6$\sigma$ for \feh) and that have two consecutive radial orders that match (any) two observed radial orders within 4\%. The latter implicitly sets an upper limit to the absolute value of the surface corrections for those modes. Then, a fine interpolation is carried out along each pruned track such that the changes in all frequencies within the observed range vary less than the typical observed error. 

We assigned a prior probability to each model formed by the product of the timespan each model represents in the track and a weight from assuming a Salpeter initial mass function. We assigned a constant prior to metallicity and star formation rate. Then, to compute the likelihood of each model, we used the reduced $\chi^2_r$, computed as:
\begin{equation}
\chi^2_r = \frac{1}{N_{\rm obs}-N_{\rm dof}} \sum_{{\rm i}=1,N_{\rm obs}}{\left(\frac{Q_{\rm mod,i}-Q_{\rm obs,i}}{\sigma_{\rm obs,i}}\right)^2}, 
\end{equation}
where the first sum extends over \teff, \feh, and \lstar, and the oscillation modes (Table~\ref{tab:freqs}). Therefore, $N_{\rm obs}$=52. Model frequencies were corrected by surface effects using the two-parameter correction introduced by \citet{ball_new_2014}. The number of degrees of freedom in the models is therefore $N_{\rm dof}$=5 (age, mass, [Fe/H] and the two parameters of the surface correction). The likelihood is then proportional to $\exp{(-\chi^2_r/2)}$. Following these steps, each model was then assigned a probability, which was the product of the prior and likelihood, and probability distribution functions were built for each quantity of interest. 

\subsubsection{MESA Model 1} 
M.J.\ used the Modules for Experiments in Stellar Astrophysics (MESA; \citealt{paxton_modules_2011, paxton_modules_2013, paxton_modules_2015, paxton_modules_2018, paxton_modules_2019}) software suite version 12778 and the GYRE stellar oscillation program \citep{townsend_gyre_2013} version 5.2. 
The physical assumptions adopted in the evolutionary calculations include the \citet{asplund_chemical_2009} opacities and solar mixture, the \texttt{photosphere\_tables} atmospheric boundary conditions, the \texttt{`Henyey`} mixing length scheme \citep{Henyey1965}, the \texttt{`basic`} nuclear network option, which includes all elements and isotopes necessary for low-mass burning chains up through the helium flash, and the \texttt{`exponential`} treatment of convective overshooting, with $f_\text{ovs}= 0.0014$ times the pressure scale height and primary convective coefficient $f_0=0.002$.

Three different fitting schemes for the asteroseismic data were considered: (A) fitting to all modes, including $\ell=1$; (B) fitting only to $\ell = 0$ and $2$ modes; and (C) fitting only to $\ell = 0$ modes. The rank orderings of best-fitting models as determined by methods (B) and (C) were the same.  
In all cases, the surface corrections of \citet{BallGizon2014} were applied to the frequency fitting, with additional seismic features (e.g. $\nu_\text{max},\Delta \nu$ ) calculated according to \citet{White2011}.

The fits to individual modes and classical constraints were combined and weighted according to the methods described in \citet{Joyce2018b} and \citet{MurphyJoyce2021}. 

\subsubsection{MESA Model 2} 
D.S.\ used the astero extension in MESA (v7503) \citep{paxton_modules_2011,paxton_modules_2013} with the default settings for the model physics, except for the following. We used \citet{asplund_chemical_2009} solar abundances, the 'o18\_and\_ne22' nuclear net, the 'Kunz' C12 rates, and 'jina reaclib' N14 rates.  For the outer boundary condition we used the 'photosphere\_tables'. 
For the fitting procedure the non-seismic input observables were effective temperature, luminosity, and metallicity corrected for the star's high alpha element content using the formulation by \citet{salaris_alpha_1993}. The seismic input was the radial mode frequencies. During model fitting and optimisation, the non-seismic input was weighted 1/3 with 2/3 to the seismic input. We corrected the model frequencies using the 'cubic' formulation by \citet{ball_new_2014}, but found no significant difference when using the 'combined' formulation. We set mass, metallicity, and the mixing length parameter as free parameters, while keeping overshoot fixed at 0.015 (exponential prescription). We used an initial helium fraction of 0.248 and an enrichment ratio of 1.4.

\subsubsection{MESA Model 3} 
Y.L.\ used MESA (version r15140) and GYRE (version 5.2), adhering mostly to the default settings, except for specific deviations outlined below. The metal mixture AGSS09 and its corresponding opacities were employed in the models. The Henyey mixing length approach \citep{Henyey1965} was used for the prescription of convection. Convective overshoot was not incorporated. The atmosphere boundary condition was set to \texttt{photosphere tables}, which are pre-calculated model atmosphere tables. A grid of stellar models were calculated with initial helium abundance, stellar mass, metallicity, and the mixing length parameter treated as free parameters. These models were constructed with frequencies and corrected with the surface effect in line with the inverse-cubic method proposed by \citet{BallGizon2014}. All models were fitted to the oscillation modes and the classical constraints, including $L$, $\teff{}$, and $\feh{}$.

\subsubsection{MESA Model 4} 
J.M.J.O. constructed MESA models using MESA r12778 and GYRE v6.0. This set of models was constructed with diffusion and settling of helium and metals, an Eddington-gray atmospheric boundary condition, and only a small amount ($f_\mathrm{ov} = 0.001; f_0=0.0005$) of exponential convective overshoot for numerical conditioning. Unlike the other MESA modelling runs, this modelling was also performed with the stellar $\mathrm{[Fe/H]}$ computed with respect to the solar chemical abundances of \cite{grevesse_standard_1998}, with further initial enrichment to match the observational value of $[\alpha/\mathrm{Fe}]$. Correspondingly, OPAL opacity tables appropriate to the GS98 chemical mixture, computed with respect to an alpha-enhancement of $[\alpha/\mathrm{Fe}] = +0.4$, were also adopted. Accordingly, the prescription of \cite{salaris_alpha_1993} was \emph{not} used; rather, the actual enriched surface values of $\mathrm{[Fe/H]}$ in these models were constrained directly against the observational measurement.

Only modes with confident identifications of $\ell$ were included in the asteroseismic constraint. Radial p-modes were computed using the surface boundary conditions of \cite{christensen-dalsgaard_adiplsaarhus_2008}, and nonradial pure p- and g-mode frequencies, and associated coupling matrices, were computed using the $\pi/\gamma$ isolation scheme of \cite{ong_semianalytic_2020}. The radial and quadrupole p-modes were corrected using the two-term prescription of \cite{ball_new_2014}. For the dipole modes, a nonlinear surface term correction was applied (the generalisation to the \citealt{ball_new_2014} correction described in \citealt{ong_mixed_2021}), in order to fully account for near-degeneracy avoided crossings between the p- and g-modes, using the same coefficients $a_{-1}$ and $a_3$ as fitted against the even-degree modes. To avoid having to perform the eigendecomposition of a large matrix for every evaluation of the seismic likelihood function, the effective frequency-dependent asymptotic coupling strength $q(\nu)$ was calculated using the expression derived in \cite{ong_mode_2023}, permitting mixed-mode frequencies to be computed as the roots of the asymptotic eigenvalue equation of e.g. \cite{unno_nonradial_1979}.

The total log-likelihood function we adopted for this exercise was the equally-weighted sum of $\chi^2$ statistics for the observed metallicity, effective temperature, and the luminosity, as well as the reduced $\chi^2$ statistic for the surface-corrected mode frequencies. Maximum-likelihood point estimates were initially derived by minimizing the negative log-likelihood using the differential-evolution algorithm, implemented in \cite{yabox}, varying the initial mass, helium mass fraction, metal fraction, and mixing length parameter $\alpha_\text{MLT}$ as optimisation parameters. Every stellar model constructed along the optimisation trajectory was retained, effectively constituting a nonuniformly sampled grid of stellar models. We then estimate the marginal posterior median and $\pm1\sigma$ quantiles of the fundamental properties as weighted averages with respect to this grid of models, with the weights being both proportional to the likelihood function, and inversely proportional to the sampling function of the grid (estimated with a kernel density estimator).

\subsubsection{BASTA}
V.A.B.-K.\ used the BAyesian STellar Algorithm (BASTA) code with a grid of models computed using Garstec with input physics as described in \citet{aguirre_borsen-koch_bayesian_2022}. The grid implements $\alpha$ enhanced and depleted mixtures of the \citet{asplund_chemical_2009} solar abundances, ranging from $\afeh = -0.2$ to 0.6 in steps of 0.1\,dex. Oscillation frequencies were calculated using ADIPLS \citep{christensen-dalsgaard_adiplsaarhus_2008} with surface corrections following the cubic formulation by \citet{ball_new_2014}. Parameter inference follows a Bayesian approach as described by \citet{aguirre_borsen-koch_bayesian_2022} and used the spectroscopic constraints (\teff, \feh\ and \afeh), Gaia luminosity and radial modes.

\begin{table*}
\begin{center}
\caption{Observed and Derived Properties for \target}\label{tab:stellar}
\begin{tabular}{l c c c c c c c}
\tableline\tableline
\noalign{\smallskip}
KIC ID \& \kep\ Magnitude & \multicolumn{3}{c}{8144907} & \multicolumn{2}{c}{$Kp=13.1$} & \multicolumn{2}{c}{\citet{brown_kepler_2011}}  \\
Gaia DR3 ID \& $G$ Magnitude & \multicolumn{3}{c}{2105367821469848704} & \multicolumn{2}{c}{$G=12.86$} & \multicolumn{2}{c}{\citet{lindegren_gaia_2021}}  \\
2MASS ID \& $K$ magnitude & \multicolumn{3}{c}{J18485977+4401183} & \multicolumn{2}{c}{$K=11.11$} & \multicolumn{2}{c}{\citet{skrutskie_two_2006}}  \\
\hline
\hline
Bol.\ Flux ($10^{-10}$\,erg\,s$^{-1}$\,cm$^{-2}$) & \multicolumn{5}{c}{$2.023 \pm 0.057$} & \multicolumn{2}{c}{this work}  \\
Luminosity $L$ ($\lsun$) &  \multicolumn{5}{c}{$11.13\pm0.45$} & \multicolumn{2}{c}{this work}  \\
\hline
Freq.\ of max.\ power \numax (\muHz) & \multicolumn{5}{c}{$208.2\pm1.1$} & \multicolumn{2}{c}{\citet{yu_asteroseismology_2018}} \\
Large Separation \dnu (\muHz) & \multicolumn{5}{c}{$17.47\pm0.092$} & \multicolumn{2}{c}{\citet{yu_asteroseismology_2018}} \\
Period Spacing $\Delta \Pi$ (sec) & \multicolumn{5}{c}{$84.83\pm$0.02} & \multicolumn{2}{c}{\citet{kuszlewicz_mixed-mode_2023}} \\
\hline
Eff.\ Temperature \teff\, (K) &  \multicolumn{5}{c}{$5400\pm200$} & \multicolumn{2}{c}{this work}  \\
Iron Abundance \feh (dex) & \multicolumn{5}{c}{$-2.66\pm0.08$} & \multicolumn{2}{c}{this work}  \\
Alpha Abundance \afeh (dex) & \multicolumn{5}{c}{$0.38\pm0.06$} & \multicolumn{2}{c}{this work}  \\
\hline
\hline
Model & BeSPP  & MESA1 & MESA2 & MESA3 & MESA4 & BASTA & Adopted \\
\noalign{\smallskip}
\hline
Mass, \mstar\ (\msun)           & 0.794  & 0.792  & 0.800  & 0.786  & 0.774  & 0.809 & \mass         \\
Radius, \rstar\ (\rsun)         & 3.617  & 3.625  & 3.620  & 3.647  & 3.572  & 3.630 & \radius      \\
Luminosity, \lstar\ (\lsun)     & 10.42  & 11.96  & 12.61  & 11.81  &  11.06  & 10.79 & \lum      \\
Density, \rhostar\ ($10^{2} \rhosun$)  & 2.37 & 2.34 & 2.38 & 2.28& 2.39   & 2.39& \density      \\
Surface Gravity, \logg\ (cgs)   & 3.221  & 3.218  & 3.223  & 3.210  & 3.221  & 3.226 & \loggval     \\
Age, t (Gyr)                    & 11.96  & 12.40  & 12.25  & 12.76  & 12.68  & 11.81 & \age  \\
\tableline\tableline
\noalign{\smallskip}
\end{tabular}
\label{tab:results}
\end{center}
\end{table*}

\subsubsection{Summary}
\label{sec:summary}
Table \ref{tab:results} summarizes the properties for \target. We observe excellent agreement between modeling methods, with a scatter of $\approx$2--3\% in mass and age. Some methods reported results excluding and including the Gaia luminosity, which were largely consistent, demonstrating that there is no strong disagreement between the luminosities from asteroseismic constraints and Gaia. We adopt self-consistent values from BeSSP, which reported a solution close to the median mass over all methods.  For uncertainties, we quote random errors using the mean uncertainty of all methods that fitted non-radial modes and systematic errors as the standard deviation over all methods. Remarkably, the systematic uncertainties for mass and age are a factor of $\approx$\,1.5-2 smaller than the random uncertainties.  

Given the similarity in methods, our quoted systematic errors only capture part of the uncertainties. Possible additional systematic errors may arise from mass loss. However, this is not expected to be above our mass uncertainty budget for a star near the base of the RGB \citep{schroder_new_2005, mullan_mass_2019}. In principle, additional errors may arise from surface corrections: for example, the \citet{ball_new_2014} correction was shown to induce large amounts of dispersive systematic errors in inferred stellar properties when using even the most p-dominated modes, which have inertiae that are not representative of the underlying pure p-modes \citep{ong_differential_2021, ong_mixed_2021}. However, the generalised prescription of \citet{ong_mixed_2021-1}, which we apply in Section 3.3.5, accounts for nonlinear mixed-mode coupling, and yields results that are consistent with other methods applying the standard \citep{ball_new_2014} correction. 
We therefore conclude that surface correction method does not induce a major systematic error in our derived mass and age for \target. Additional estimates for contributions to the error budget for similar stars have been computed in, e.g., \citet{silva_aguirre_aarhus_2020, christensen-dalsgaard_aarhus_2020, tayar_guide_2022, Joyce2023, YingChaboyer2023}.

\section{Implications for Models of Very Metal-Poor Stars}

Metal-poor stars are important testbeds for stellar models. Input physics such as opacities, interior mixing and convective energy transport prescriptions depend on the chemical composition of a star, and become more uncertain for non-solar metallicities \citep[e.g.][]{JoyceChaboyer2018NotAllStars}. The fidelity of such models has a wide impact on astrophysics, including our understanding of galaxies through simple stellar populations models \citep{conroy_propagation_2009}.

Our results demonstrate an excellent match between observations and models, including for mixed dipole modes which probe the structure near the core. The right panel of Figure \ref{fig:seismo_obs} compares the best-fitting BeSSP model to the observed oscillating frequencies. The model frequencies agree on average with the observations to within $\approx$\,0.06\muHz\ ($\approx$\,0.03\%), comparable to modeling results for solar-metallicity stars \citep[e.g.][]{silva_aguirre_ages_2015}. The models also provide a good match to the atmospheric constraints, supporting the fidelity of non-LTE corrections \citep[e.g.][]{bergemann_non-lte_2012}, although our \teff\ uncertainty (200\,K) is conservative. Overall, the results imply that state-of-the-art stellar models with standard physical assumptions 
can reliably reproduce detailed observations of the interior and the atmosphere of a very metal-poor star.

\section{Asteroseismic Scaling Relations}

Asteroseismic scaling relations use the global oscillation parameters \numax\ and \dnu\ to calculate fundamental properties by scaling from the Sun. The relations are widely used in the era of ensemble asteroseismology, including to calibrate spectroscopic tracers of stellar mass, which are then used to infer ages with large surveys \citep{Martig2016,ho_masses_2017,ness_spectroscopic_2016}. Testing scaling relations has become a major subfield of asteroseismology \citep{huber_fundamental_2012, gaulme_testing_2016,sahlholdt_asteroseismic_2018,themesl_oscillating_2018,zinn_testing_2019,hekker2020scaling,liyg2021scaling}. 

Early results revealed strong inconsistencies between masses from scaling relations and astrophysical expectations for metal-poor stars \citep{epstein_testing_2014}. Several corrections to scaling relations have been proposed \citep{white_asteroseismic_2011,viani_investigating_2018}. However, all fundamental calibrators to date (such as stars in eclipsing binaries, measured angular diameters, or masses from frequency modeling) have near-solar metallicity. \target\ provides an opportunity to test scaling relations in the very-metal poor regime.

Using \numax\ and \dnu\ from \citet{yu_asteroseismology_2018}, our spectroscopic \teff, and solar reference values from \citet{huber_solar-like_2011} we calculate a scaling relation mass of $0.98\pm0.06$\msun, which is $\approx$\,24\% larger than our adopted mass from individual frequency modeling. Using the corrections of the \dnu{} scaling relations \citep{sharma_asfgrid_2016} calculated by \citet{stello_extension_2022} for low-metallicity stars, we obtained $0.99\pm0.06$\msun, even higher than the uncorrected values. Recent revisions on the calculated \dnu{} correction factors due to the surface terms can at most decrease the scaling mass by $\approx$8\% \citep{litd2022scaling,li_prescription_2023}, which is still unable to fully explain the difference. This supports a break down of the \numax{} scaling relation in the low-metallicity regime \citep{belkacem_underlying_2011,zhou++2023-numax} and is in tension with the suggestion that temperature scales are the main cause for overestimated scaling relation masses for metal-poor stars \citep{schonhut-stasik++2024-apok2}. 

\begin{figure*}
\begin{center}
\resizebox{\hsize}{!}{\includegraphics{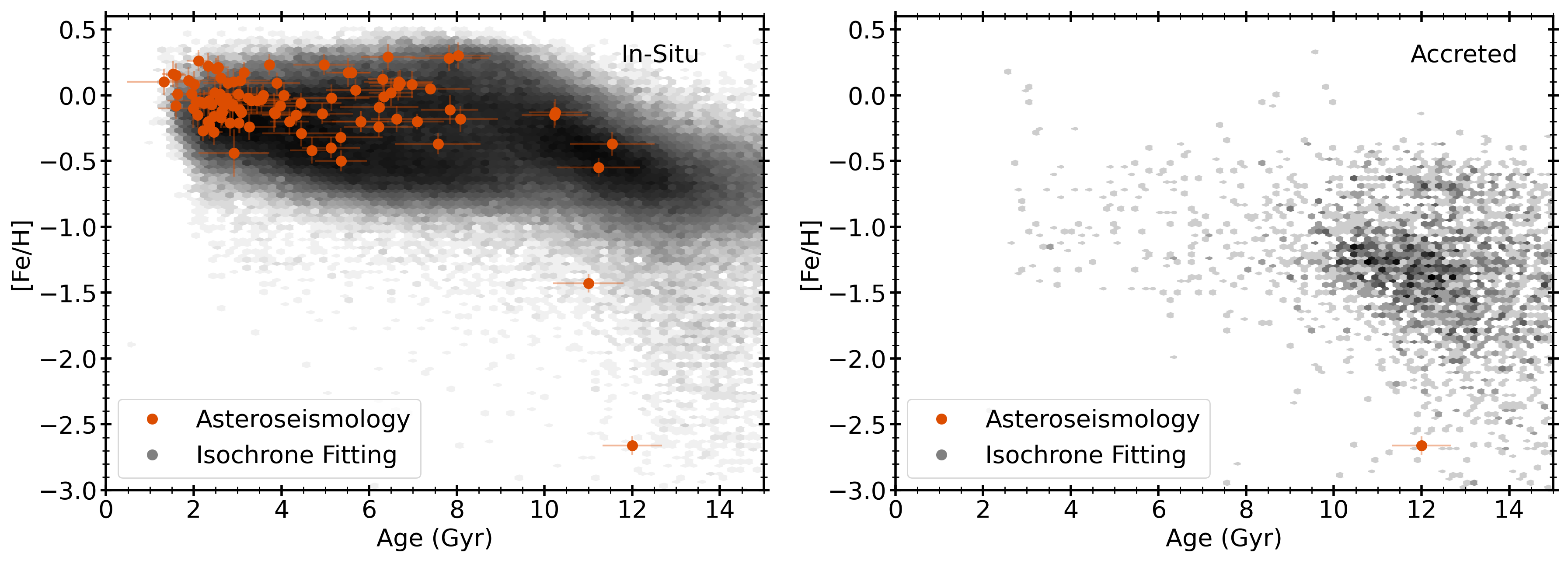}}
\caption{Left panel: Iron abundance versus age for stars with ages from detailed asteroseismic modeling (orange circles) and subgiants with ages from isochrone fitting by \citet{xiang_time-resolved_2022} (grey logarithmic heatmap) with kinematics consistent with an ``in-situ'' population ($e<0.8$, $Lz > 0$kpc\,km\,s$^{-1}$). Literature asteroseismic ages are taken from \citet{silva_aguirre_ages_2015}, \citet{silva_aguirre_standing_2017}, \citet{campante_ancient_2015}, \citet{chaplin_age_2020} and \citet{li_asteroseismology_2020}. Right panel: same as left panel but for stars with kinematics consistent with an accreted population ($e>0.8$, $Lz < -500$kpc\,km\,s$^{-1}$). The abundances and kinematics of KIC\,8144907 are consistent with either population.}
\label{fig:agefeh}
\end{center}
\end{figure*}

We can phrase this discussion in terms of the \numax\ scaling relation, in which \numax\ is approximately proportional to $g/\sqrt{\teff}$ \citep{brown_detection_1991, Kjeldsen1995, belkacem_underlying_2011}:
\begin{equation}\label{eq:fnumax}
\frac{\nu_{\rm max}}{\nu_{\rm max,\odot}} = f_{\numax} \left(\frac{M}{M_{\odot}}\right) \left(\frac{R}{R_{\odot}}\right)^{-2} \left(\frac{\teff}{T_{\rm eff,\odot}}\right)^{-1/2}.
\end{equation}
The factor $f_{\numax}$ is used to quantify the deviation from this scaling relation \citep{sharma_asfgrid_2016, li_realistic_2024}. Because our models did not use \numax\ as an input, we can use the properties of the best-fitting model to test the departure from the \numax\ scaling relation. We do this by calculating $f_{\numax}$ from the mass, radius and effective temperature of the model, and the measured value of \numax, which gives $f_{\numax}=1.056$. This confirms the trend found by \citet{li_realistic_2024} that $f_{\numax}$ increases towards lower metallicity. \rev{It is in contrast to \citet{viani_investigating_2018}, who investigated the acoustic cutoff frequency and found $f_{\numax}$ lower than 1 in metal-poor stars. This disagreement is not well understood. }

We also considered the new method of defining and measuring \numax\ introduced by \citet{sreenivas_2024}. Using their nuSYD pipeline, they found a value of \numax\ for \target\ (after correcting for bias) of $209.0 \pm 0.9\,\muHz$ which, combined with their solar value of 3047\,\muHz, means \numax\ is 0.0686 times solar.  This value is 2\% higher than 0.0674 times solar found by \citet{yu_asteroseismology_2018}, increasing the mass from the scaling relations by 6\% and making the disagreement with the best-fit model slightly greater ($f_{\numax} = 1.075$).

\section{An Asteroseismic Age for a Very Metal Poor Star}

\target\ is consistent with the canonical definition of a local Galactic halo star based on its metallicity ($\feh < -1$) and kinematics ($|V-V_{\rm{LSR}}|>200$\,km\,s$^{-1}$). Gaia led to the discovery that many local stars with halo-like kinematics formed in the Milky Way disc but were later kinematically heated through merger events \citep{bonaca_gaia_2017}. At fixed metallicity, such ``in-situ'' halo stars are generally more enriched in $\alpha$ elements compared to stars that have been accreted from dwarf galaxies that merged with the Milky Way, such as Gaia-Enceladus \citep{helmi_merger_2018,belokurov_ges_2018}. These low and high $\alpha$ sequences merge for very metal-poor stars, which are rare in Gaia, making the distinction of their origin based on kinematics and chemical abundances more difficult. In general, most stars with $\feh < -1$ are expected to have been accreted from satellite galaxies \citep{di_matteo_milky_2019}. However, results for high-latitude halo stars have suggested that the oldest 
in-situ stars can extend to at least $\feh \approx -2.5$\,dex \citep{carollo_evidence_2019, carter_ancient_2021, conroy_birth_2022}, and some very metal-poor stars have also been associated with the oldest stars of the thin disc \citep{nepal_discovery_2024}. 

To investigate the dynamical origin of \target, we compute its orbital kinematics using \texttt{galpy} \citep{galpy} assuming a left-handed Galactocentric coordinate frame with $[V_{R, \odot}, V_{\phi, \odot}, V_{Z,\odot}] = [-12.9, 245.6, 7.78]\,$km/s \citep{Drimmel_2018}, $Z_{\odot}=20.8\,$pc \citep{Bennett_2019}, and $R_0=8.122\,$kpc \citep{GRAVITY_2018}. The star's orbit is integrated using time-steps of 1 Myr up to a total integration time of 20 Gyr assuming the \texttt{McMillan2017} potential \citep{McMillan_2017}, which reveals that the star follows an eccentric ($e=0.75$) orbit with a slight retrograde motion relative to the Milky Way's rotation ($L_z=-0.5 \times 10^{3}$\,kpc\,km/s). Following \citet{naidu_evidence_2020}, these chemo-dynamical properties make \target\ consistent with either the metal-weak thick disc or a Gaia-Enceladus member. 
Following \citet{feuillet20}, \target\ would be classified as a possible Gaia Enceladus member, although its metallicity would be unusually low for that population. The properties of \target\ also appear consistent with the Atari disk \citep{mardini_atari_2022}, which is suspected to have an accretion origin. \rev{The catalog by \citet{dodd_gaia_2023}, based on clustering algorithm by \citet{ruiz-lara_substructure_2022}, lists \target\ as not being associated with dynamical substructures resulting from merger debris.}

Irrespective of origin, an asteroseismic age for a very metal poor star provides a valuable anchor point for formation events in our galaxy. Figure \ref{fig:agefeh} shows the age-metallicity relation for stars with asteroseismic ages from detailed frequency modeling with uncertainties better than 1\,Gyr. We also show a larger sample of subgiant stars with ages determined from isochrone fitting \citep{xiang_time-resolved_2022}. We divided the sample into likely ``in-situ'' and ``accreted'' stars based on eccentricity and angular momentum following \citet{conroy_birth_2022}. While this selection is not exact and turn-off ages can be affected by systematic errors due to stellar multiplicity, these cuts give broadly representative distributions. We observe that \target\ sits at the young edge for both distributions at its metallicity. If \target\ \rev{was formed in-situ and thus belongs the earliest phase of the galaxy}, 
its age would imply that substantial star formation in the Milky Way that led to the high-$\alpha$ thick disc did not commence until $\approx$\,12\,Gyr ago (redshift $z\approx$\,3). If \target\ was accreted (e.g., from Gaia Enceladus), it would demonstrate that the Milky Way has undergone merger events for at at least $\approx$\,12\,Gyr. No star with detailed asteroseismic frequency modeling has yet been found to be substantially older than 13\,Gyr.

\section{Conclusions}

We presented the discovery of the most metal-poor star to date for which detailed asteroseismic modeling is possible ($\feh = \met$ and $\afeh= \alp$). Our main conclusions are as follows:

\begin{itemize}

\item We demonstrate that state-of-the-art stellar structure and atmosphere models with non-LTE corrections can successfully reproduce observations of a very metal-poor star, including a large number of oscillation frequencies that probe the stellar core. Masses and ages derived from different methods agree to within $\approx$2-3\%, with random errors exceeding systematic errors by a factor of $\approx$\,1.5-2. This implies that such stellar models make reliable predictions for very metal-poor stellar populations, although we note that not all possible sources of systematic errors were covered in the applied methods.

\item We measure a mass of \mass\ \msun, and confirm that asteroseismic scaling relations overestimate stellar masses by $\approx$\,20\% for a very metal-poor star. Corrections to the \dnu\ relation do not improve the comparison, and our results are consistent with a metallicity dependence for corrections to the \numax\ relation. The results demonstrate that large-scale asteroseismic results based on average seismic parameters (and calibrations that are based on such samples) have to be used with caution for stars with $\feh<-1$.

\item We measure an age of \age\ Gyr, the first such estimate from detailed asteroseismic modeling for a very metal-poor star. The chemical abundances and kinematics of the star are consistent with either an \rev{ancient in-situ star} (implying that substantial star formation in the Galaxy did not commence until $z\approx$\,3) or with being accreted from a dwarf satellite such as Gaia Enceladus (implying that Milky Way has undergone merger events for at least 12\,Gyr).

\end{itemize}

\target\ demonstrates the value of systematic spectroscopic follow-up for stars with asteroseismic detections from space-based telescopes, in particular those with high S/N detections from \kep\ and in the TESS continuous viewing zones. Accumulating more examples will be critical for the success of Galactic archeology, in particular for improving the reliability spectroscopic age scales that are calibrated on asteroseismology \citep[e.g.][]{Martig2016}. Given the large number of projected asteroseismic detections from TESS \citep[e.g.]{hon_quick_2021}, this will require large-scale spectroscopic surveys such as SDSS-V \citep{kollmeier_sdss-v_2017}, 4MOST \citep{de_jong_4most_2012}, WEAVE \citep{dalton_weave_2012}  MSE \citep{the_mse_science_team_detailed_2019,bergemann_stellar_2019} and WST \citep{bacon_wst_2024}.

Data and code to reproduce all figures in this paper can be found at \url{https://github.com/danxhuber/very-metal-poor-seismology}. The \kep\ data used in this paper can be found in MAST: \dataset[10.17909/xsvn-fe82]{http://dx.doi.org/10.17909/xsvn-fe82}.

\vspace{0.5cm}
\noindent

The authors wish to recognize and acknowledge the very significant cultural role and reverence that the summit of Maunakea has always had within the Native Hawaiian community. We are most fortunate to have the opportunity to conduct observations from this mountain.

Some of the data presented herein were obtained at Keck Observatory, which is a private 501(c)3 non-profit organization operated as a scientific partnership among the California Institute of Technology, the University of California, and the National Aeronautics and Space Administration. The Observatory was made possible by the generous financial support of the W. M. Keck Foundation. This research is based in part on data collected at the Subaru Telescope, which is operated by the National Astronomical Observatory of Japan. 

D.H. acknowledges support from the Alfred P. Sloan Foundation, the National Aeronautics and Space Administration (80NSSC19K0597), and the Australian Research Council (FT200100871). A.S. acknowledges grants Spanish program Unidad de Excelencia Mar\'{i}a de Maeztu CEX2020-001058-M, 2021-SGR-1526 (Generalitat de Catalunya), and support from ChETEC-INFRA (EU project no. 101008324). D.S. is supported by the Australian Research Council (DP190100666). E.N.K.\ acknowledges support from NSF CAREER grant AST-2233781.
T.R.B. acknowledges support from the Australian Research Council through Laureate Fellowship FL220100117.
M.J. gratefully acknowledges funding of MATISSE: \textit{Measuring Ages Through Isochrones, Seismology, and Stellar Evolution}, awarded through the European 
Commission's Widening Fellowship.  
This project has received funding from the European Union's Horizon 2020 research and innovation programme.
A.A.P. acknowledges the support by the State of Hesse within the Research Cluster ELEMENTS (Project ID 500/10.006).
This research was supported by the Munich Institute for Astro-, Particle and BioPhysics (MIAPbP) which is funded by the Deutsche Forschungsgemeinschaft (DFG, German Research Foundation) under Germany´s Excellence Strategy – EXC-2094 – 390783311.

\software{This research made use of echelle \citep{hey_echelle_2020}, isoclassify \citep{huber_asteroseismology_2017, berger_gaiakepler_2020}, Lightkurve \citep{lightkurve_collaboration_lightkurve_2018}, Matplotlib \citep{hunter_matplotlib_2007}, numpy \citep{harris_array_2020} and scipy \citep{virtanen_scipy_2020}.}

\facilities{UH:2.2m, Keck, Subaru, Kepler}

\appendix

\section{Metal Poor Star Catalog}

\begin{table}[h!]
\begin{center}
\caption{Metal poor oscillating Kepler red giants}
\vspace{0.1cm}
\begin{tabular}{c c c}        
KIC & \numax & \feh  \\
\hline       
5522939 & 22.1 & -1.3 \\
5938297 & 70.5 & -2.7 \\
6278241 & 14.7 & -1.5 \\
7259558 & 24.2 & -1.3 \\
7635173 & 30.7 & -1.2 \\
7660280 & 36.7 & -1.3 \\
7879709 & 18.4 & -1.7 \\
8411296 & 11.3 & -1.6 \\
8611078 & 58.7 & -1.6 \\
9508233 & 39.0 & -1.1 \\
10728519 & 84.9 & -1.1 \\
11445766 & 17.8 & -1.4 \\
11460115 & 105.5 & -1.3 \\
11700922 & 4.8 & -2.1 \\
11704816 & 92.9 & -1.8 \\
\hline 
\end{tabular}
\end{center}
\label{tab:app}
\flushleft Notes: Other stars found to be metal-poor ($\feh < -1$) by our spectroscopic survey. \numax\ values are taken from \citet{yu_asteroseismology_2018}, \feh\ values were derived using iSpec. Uncertainties on \feh\ are $\approx$\,0.1\,dex based on a comparison to metallicities derived from APOGEE for a subset of stars. 
\end{table}

\bibliography{refs/references,refs/otherrefs, refs/Joyce_refs, refs/ditte_refs, refs/yaguang_refs, refs/ong_refs}{}
\bibliographystyle{aasjournal}

\suppressAffiliationsfalse
\allauthors

\end{document}